\documentclass[aps, floatfix, twocolumn, superscriptaddress]{revtex4-2}

\usepackage{bm}
\usepackage{amsmath}
\usepackage{mathtools}
\mathtoolsset{showonlyrefs=true,showmanualtags=true}
\usepackage{amsfonts}
\usepackage{siunitx}
\usepackage{graphicx}

\newcommand{\be}{\begin{equation}}
\newcommand{\ee}{\end{equation}}

\usepackage{hyperref}

\begin{document}

\title{Rotating Spin Wave Modes in Nanoscale M\"obius Strips}

\author{Ashfaque Thonikkadavan}
\affiliation{Universit{\'e} de Strasbourg, CNRS, Institut de Physique et Chimie des Mat{\'e}riaux de Strasbourg, F-67000 Strasbourg, France}

\author{Massimiliano d'Aquino}
\email{mdaquino@unina.it}
\affiliation{Department of Electrical Engineering and ICT, University of Naples Federico II, Naples, Italy}

\author{Riccardo Hertel}
\email{riccardo.hertel@ipcms.unistra.fr}
\affiliation{Universit{\'e} de Strasbourg, CNRS, Institut de Physique et Chimie des Mat{\'e}riaux de Strasbourg, F-67000 Strasbourg, France}

\begin{abstract}
Curved and topologically nontrivial magnetic structures offer new pathways to control spin-wave behavior beyond planar geometries. Here, we study spin-wave dynamics in Möbius-shaped soft-magnetic nanostrips using micromagnetic simulations. By comparing single-, double-, and triple-twisted Möbius strips to a topologically trivial bent ring, we isolate the roles of helical twist and non-orientable topology. Möbius geometries exhibit non-degenerate mode doublets associated with counterpropagating spin waves, in contrast to the standing-wave doublets in the trivial case. This splitting arises from a twist-induced geometric (Berry) phase that breaks propagation symmetry, producing non-reciprocal dispersion relations. The Möbius topology further imposes antisymmetric boundary conditions, resulting in half-integer wavelength quantization. Local RF excitation allows for the selective generation of spin waves with defined frequency and direction. An analytical model reproduces the dispersion behavior with excellent agreement. These results highlight how geometric and topological design can be leveraged to engineer spin-wave transport in three-dimensional magnonic systems.
\end{abstract}
\maketitle

\section{Introduction}
The emergence of three-dimensional (3D) nanomagnetism has opened up new frontiers in the study of magnetization dynamics, extending beyond the well-charted territory of planar thin-film systems~\cite{fernandez-pacheco_three-dimensional_2017,gubbiotti_three-dimensional_2019,gubbiotti_2025}. While magnetic thin films have been intensely investigated for decades, the ability to fabricate and probe 3D magnetic architectures~\cite{fernandez-pacheco_writing_2020,fischer_launching_2020, donnelly_element-specific_2015,sanz-hernandez_artificial_2020,skoric_layer-by-layer_2020,williams_two-photon_2018,keller_direct-write_2018,teresa_review_2016} enables the exploration of physical phenomena that have no analogue in planar systems, where three-dimensional geometry, topology, and curvature can profoundly modify magnetic textures and spin-wave dynamics~\cite{hertel_curvature-induced_2013, 
gaididei_curvature_2014,
otalora_curvature-induced_2016,
streubel_magnetism_2016,
heyderman_mesoscopic_2021,
sheka_fundamentals_2022}.

Within this broader context, 3D magnonics~\cite{kruglyak_magnonics_2010,gubbiotti_three-dimensional_2019,chumak_roadmap_2022} has rapidly become a vibrant research field, seeking to understand and harness spin-wave phenomena in complex 3D geometries~\cite{gubbiotti_2025}. Parallel to this, curvilinear magnetism~\cite{makarov_curvilinear_2022}, where curvature and torsion act as effective interactions, has gained growing attention for the novel magnetic textures and dynamic responses it enables.

Due to their unique geometry and topology, Möbius ring structures have attracted interest across multiple areas of physics involving transport and oscillatory phenomena, including photonics~\cite{wang_experimental_2023}, quantum systems~\cite{flouris_curvature-induced_2022}, phonon dynamics~\cite{nishiguchi_phonon_2018}, and microwave resonators~\cite{hamilton_absorption_2021}. 
In this work, we investigate spin-wave dynamics in Möbius-shaped soft-magnetic nanostructures, which combine three-dimensionality, curvature, and nontrivial topology. Möbius strips possess a non-orientable surface and intrinsic helical twist~\cite{guo_non-orientable_2023}, distinguishing them from topologically trivial ring structures. We use full-scale finite-element (FEM) micromagnetic simulations to model spin-wave dynamics with high accuracy in both time~\cite{hertel_tetmag_2023} and frequency~\cite{daquino_micromagnetic_2023} domains, to capture the effects of smooth curvature and topology on mode formation and propagation.

A central result is the helicity-dependent splitting, i.e., a lift of degeneracy, between clockwise and counterclockwise propagating modes---a phenomenon absent in topologically trivial strips, where such modes are locked into standing waves by symmetry. 
Remarkably, the Möbius geometry enables the selective excitation of circularly propagating spin-wave modes using a linearly oscillating magnetic field at the appropriate frequency. We numerically extract the magnon dispersion relations for Möbius strips with one, two, and three twists, which, due to the quantization by the circular boundary conditions, result in a distinct set of frequencies and wave vectors. We utilize an analytical model~\cite{daquino_nonreciprocal_2025} that reproduces the essential spin-wave dynamics in the Möbius strips. The theoretical predictions show excellent agreement with micromagnetic simulations.  These findings highlight how topology and 3D curvature can fundamentally alter magnonic behavior, offering new strategies for engineering spin-wave transport in next-generation nanoscale devices.

\section{Results}

Previous studies on magnetic Möbius strips have focused primarily on static magnetic textures~\cite{pylypovskyi_coupling_2015}. In particular, systems with perpendicular magnetic anisotropy have been shown to support topologically protected \SI{180}{\degree} Bloch-type domain walls stabilized by the strip's non-orientable geometry. Beyond the static regime, analytical work has predicted how curvature and torsion can affect spin-wave propagation in curved magnetic structures, including Möbius geometries~\cite{sheka_torsion-induced_2015, gaididei2017magnetization}. An experimental demonstration of nanoscale Möbius-shaped magnetic rings has also been reported~\cite{skoric_layer-by-layer_2020}, highlighting the feasibility of such structures for future magnonic applications.

In contrast to these earlier works, we investigate dynamic spin-wave behavior in soft-magnetic Möbius strips using full-scale micromagnetic simulations. The ground state is a topologically trivial annular vortex, and it is primarily the dynamic response, rather than the static configuration, that is affected by the strip's geometric and topological features. In particular, the perpendicular components of the dynamic magnetization are influenced by the geometry and topology.

\begin{figure}[h]
\includegraphics[width=\linewidth]{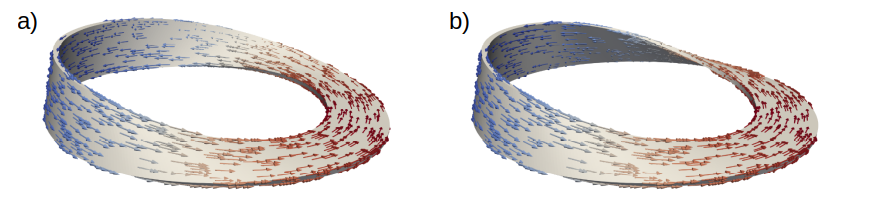}
\caption{\label{fig:geoms}Soft-magnetic annular nanostrips with identical material, geometric parameters (thickness, radius, and width), but differing topology. (a) The Möbius strip features a non-orientable surface with constant torsion along the circumference. (b) The twist-compensated ``bent'' ring is topologically trivial and has a similar local torsion, but with alternating handedness. In both cases, the equilibrium magnetic state is a macroscopic vortex configuration, with the magnetization closely aligned along the tangential direction. Spin waves propagating along the strips experience antisymmetric boundary conditions in the Möbius strip (a) and ordinary, symmetric boundary conditions in the bent ring (b).}
\end{figure}

As a model system, we consider a Permalloy Möbius strip with a diameter of \SI{100}{\nano\meter}, a width of \SI{20}{\nano\meter}, and a thickness of \SI{2}{\nano\meter}, as shown in Fig.~\ref{fig:geoms}a). Due to its nonzero thickness, the structure is technically orientable and topologically equivalent to a torus. However, in a micromagnetic sense, its strong shape anisotropy and minimal thickness constrain the magnetization to lie in-plane, with negligible variation across the thickness. As a result, the system behaves as an effectively two-dimensional magnetic structure, with its topological characteristics directly influencing the dynamic magnetization field.

We expect that key features of the Möbius geometry will strongly influence spin-wave dynamics, particularly for modes propagating along the strip's circumference. First, unlike a flat strip of identical width and thickness, the Möbius strip exhibits a continuous helical twist with an intrinsic handedness defined by its geometry. Second, its reconnection creates a non-orientable surface, leading to nontrivial periodic boundary conditions. In addition, the finite size of the structure results in a discrete, quantized set of resonance modes and frequencies, in contrast to extended planar strips.
As we will show later, these geometric and topological features produce distinct effects, including the emergence of geometric phases in the spin-wave modes. By contrast, the circular curvature introduced by closing the strip into a ring plays only a minor role.

To isolate the effects of twist and topology, we compare the Möbius strip to a control structure with the same material properties and geometric parameters (thickness, width, and radius), but differing in topology, as shown in Fig.~\ref{fig:geoms}b. Specifically, we consider a topologically trivial ``bent'' ring---similar in spirit to the photonic system studied by Wang et al.~\cite{wang_experimental_2023}---in which the strip is twisted so that its local surface is perpendicular to the ring plane on one side and parallel on the opposite side. This twist-compensated ring closely matches the M\"obius strip in local curvature over most of its length, but consists of two semicircular segments with opposite helicity and no topological twist.

This comparison allows us to disentangle the individual roles of curvature, local twist, and global topology in shaping spin-wave behavior. For brevity, we limit the discussion to the case of a left-handed Möbius strip; the right-handed version yields analogous, mirror-symmetric results. In both geometries, the equilibrium magnetic configuration (in the absence of external fields) is an annular vortex, with the magnetization lying in-plane and closely following the local tangential direction of the twisted strip.

\subsection{Resonant modes\label{eigenmodes}}
We begin by analyzing the spin wave dynamics in the case of the bent, twist-compensated ring. To characterize its intrinsic high-frequency properties, we compute the first 30 eigenmodes, covering a frequency range up to approximately \SI{45}{\giga\hertz}. Since our primary interest lies in understanding how twist and topology influence spin-wave behavior, considering only the first few modes is sufficient to analyze the system's magnonic properties. These modes exhibit well-defined azimuthal spin-wave patterns in the regime below about \SI{20}{\giga\hertz}. In contrast, at higher frequencies, the mode patterns become increasingly complex as radial standing-wave patterns occur, with wavelengths comparable to the strip width.

\begin{figure}[ht]
\includegraphics[width=\linewidth]{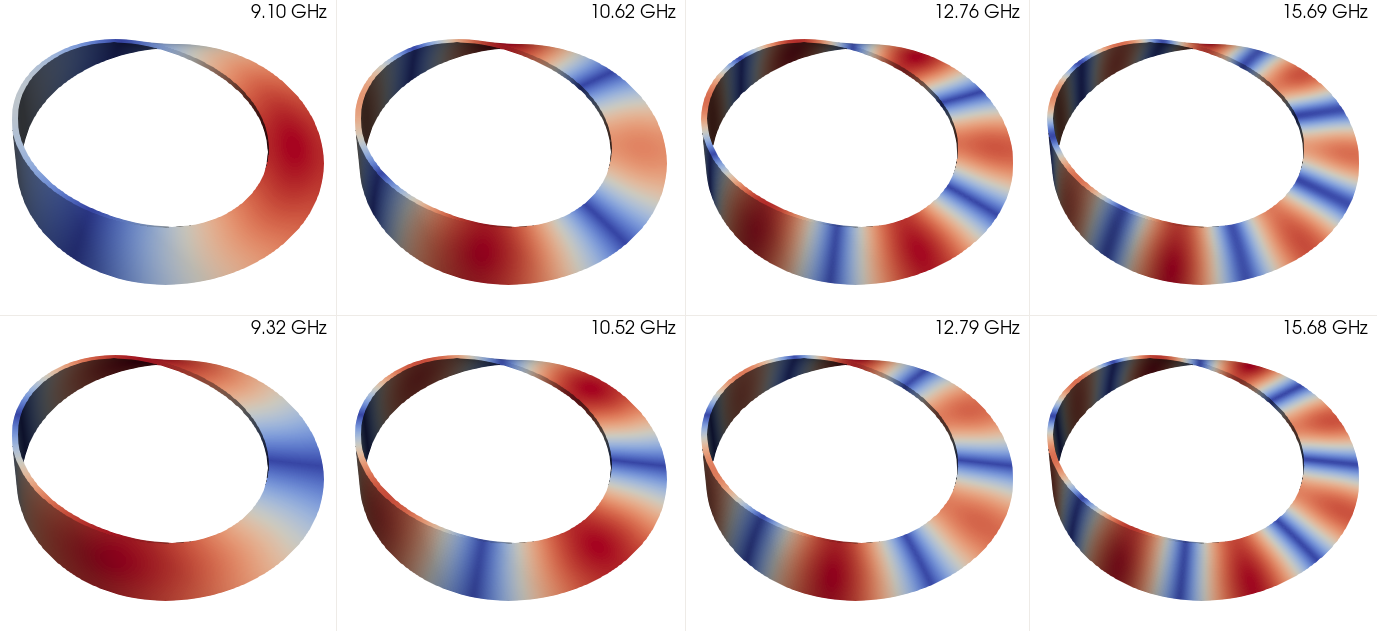}
\caption{\label{fig:modes_bent}The first eight standing-wave modes in the bent, twist-compensated ring. The color code indicates the oscillation amplitude, rescaled for each panel, with blue indicating regions of lowest amplitudes (nodes) and red regions of maximum activity (antinodes). The lowest-frequency mode, an FMR-type oscillation without a wave character at \SI{8.60}{\giga\hertz}, is omitted.}
\end{figure}

\begin{figure}[ht]
\includegraphics[width=\linewidth]{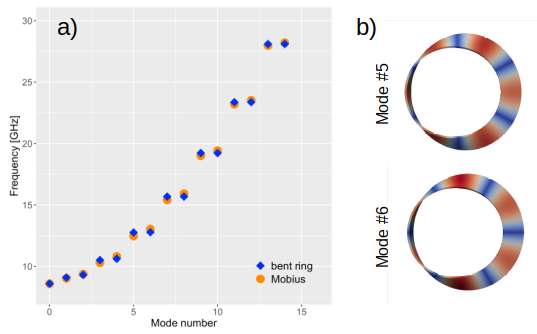}
\caption{Frequency doublets of eigenmodes in the bent ring. (a) Eigenmode frequencies as a function of mode number, starting from the lowest-frequency mode. Except for the zero-frequency mode (the ferromagnetic resonance, FMR), all modes appear as nearly degenerate pairs—frequency doublets. A similar pattern is observed in the Möbius strip. (b) Spatial profiles of an example doublet: mode \#5 and mode \#6. Both modes have the same wavelength and identical numbers of nodes and antinodes, but differ in the spatial positioning of their nodes and antinodes, which are shifted relative to each other along the ring. Color indicates oscillation amplitude, with antinodes (maximum amplitude) shown in red and nodes (minimal amplitude) in blue.\label{fig:doublets_bent}.}
\end{figure}

We observe distinct standing-wave patterns characterized by a growing number of nodes and antinodes with increasing mode frequency. These results reflect the quantization of spin-wave modes due to the finite ring circumference and its periodic boundary conditions. While such spin-wave quantization corresponds to what is expected in flat, planar magnetic rings \cite{ivanov_zaspel_jmmm2005}, we notice the appearance of frequency doublets, each corresponding to a pair of modes with the same number of nodes and antinodes, but different wave profiles, see Fig.~\ref{fig:doublets_bent}a).
The modified geometry of the bent ring leads to the appearance of mode doublets—pairs of nearly degenerate frequencies. This effect arises from the azimuthal asymmetry introduced by the bend, causing the nodes and antinodes to localize at well-defined regions of the ring. Specifically, the modes develop either maxima or minima in the area where the strip stands perpendicular or parallel to the ring plane (the inflection points of the torsion), leading to a two-fold appearance of modes of the same order.

The lowest-frequency mode, at \SI{8.60}{\giga\hertz}, is the FMR mode in which the entire sample oscillates in phase. While the phase of the oscillation is homogeneous, the oscillation amplitude varies significantly across the ring, with a clear maximum localized in the perpendicularly bent region. The following two modes, at \SI{9.09}{\giga\hertz} and \SI{9.32}{\giga\hertz}, are long-wavelength standing modes whose profile is not well-defined, as their wavelength corresponds to the ring's perimeter. Their wave profiles are difficult to interpret unambiguously as they are significantly affected by the strip's torsion, with the wave undergoing substantial changes in surface orientation over its wavelength. Nevertheless, upon close inspection, one sees that these modes represent a doublet, with one having a maximum oscillation in the perpendicular and horizontal regions and the other exhibiting maximum oscillation in the twisted branches. All subsequent modes are clearly recognizable as doublets of standing-wave patterns with an even number of (anti-)nodes. In Fig.~\ref{fig:modes_bent}, the mode doublets are grouped columnwise, with the row on the top displaying the modes where the horizontal segment on the right is a node. The bottom row shows the modes complementary to those of the top row, in which that region is an antinode.

\begin{figure}[ht]
\includegraphics[width=\linewidth]{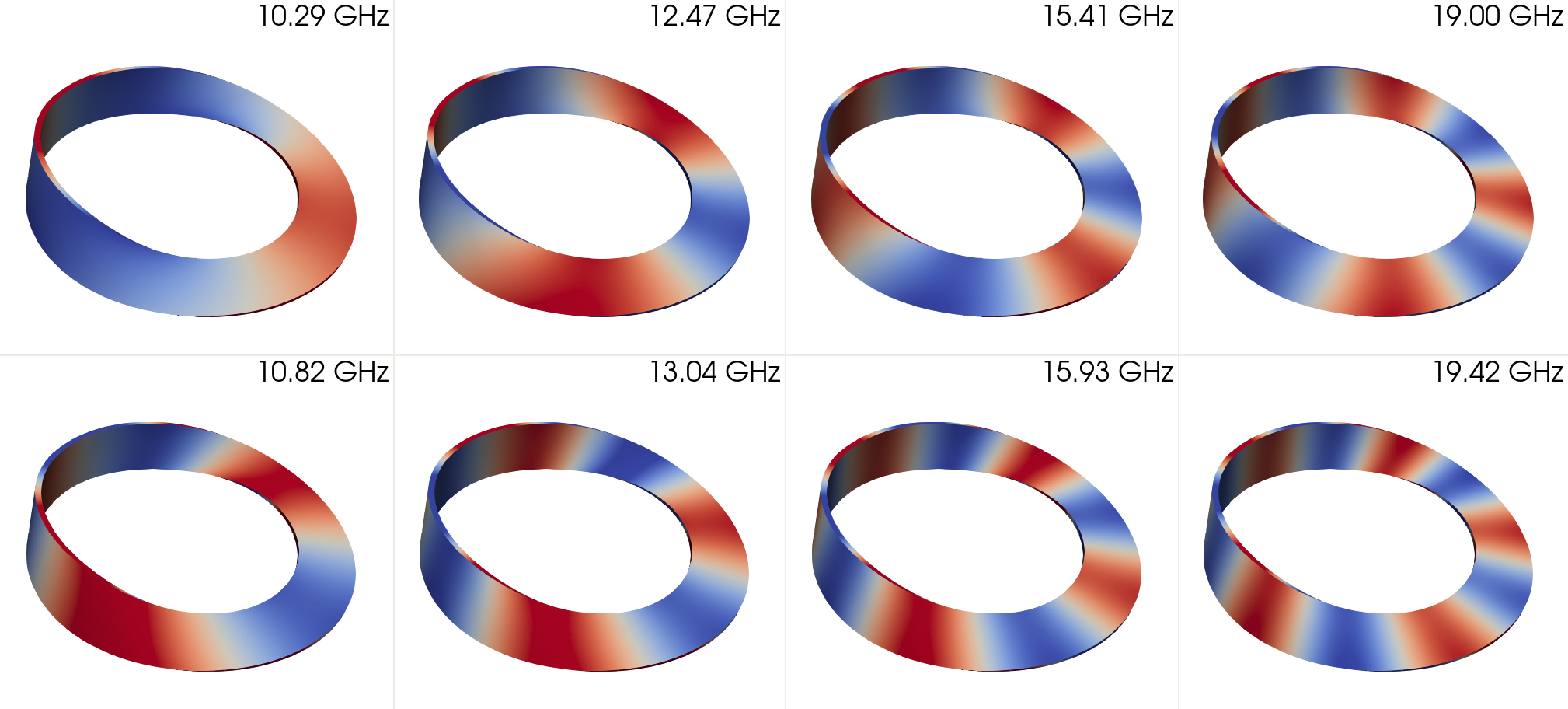}
\caption{\label{fig:mobius_wave_modes}
Spin-wave eigenmodes in the Möbius strip. Modes \#3 through \#10 are shown, with each column representing a pair of modes with similar frequency (doublets). The top row shows modes propagating clockwise; the bottom row shows the corresponding counterclockwise-propagating modes. Each image displays the mode profile at a single phase in time. Color indicates the projection of the dynamic magnetization onto the local surface normal.
}
\end{figure}

A similar appearance of frequency doublets is also observed in the eigenmodes of the Möbius strip, as shown in Fig.~\ref{fig:doublets_bent}a). In this case, too, the lowest-frequency mode is an FMR-type oscillation of uniform phase at \SI{8.60}{\giga\hertz} with a frequency nearly identical to that of the bent ring.   
Similar to the bent ring, the two next-higher frequency modes don't display a clear wave pattern, as the magnetic modulation occurs on a length scale corresponding to the ring's circumference. At higher frequencies, however, the mode character becomes clearly discernible and changes qualitatively compared to the topologically trivial ring: the standing-wave patterns are replaced by propagating modes. While these modes also form doublets, the lifted degeneracy now reflects the direction of propagation---clockwise or counterclockwise around the Möbius strip, with one circulation direction consistently showing a slightly higher frequency than the other, see Fig.~\ref{fig:mobius_wave_modes}. The modes' sense of rotation alternates strictly in the sequence of modes, revealing an apparent helicity-dependent symmetry breaking absent in the topologically trivial ring.

\begin{figure}[ht]
\includegraphics[width=\linewidth]{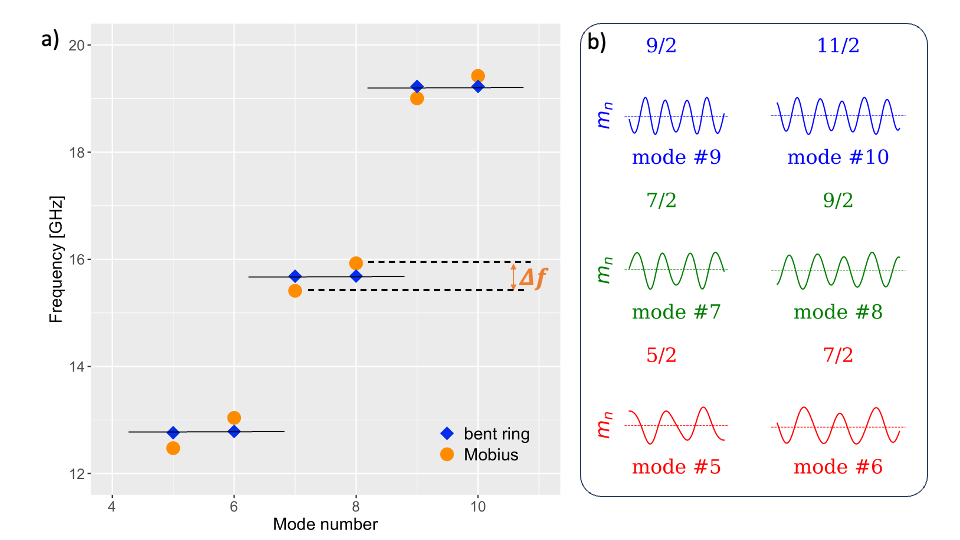}
\caption{\label{fig:doublets_mob}
Frequency splitting of spin-wave mode doublets in the Möbius strip.
(a) Mode frequencies for the Möbius strip (orange circles) and the bent ring (blue diamonds). While the bent ring exhibits nearly degenerate doublets, the Möbius strip shows a clear frequency splitting. For example, modes \#7 and \#8 in the bent ring both resonate at \SI{15.68}{\giga\hertz}, whereas the corresponding Möbius modes occur at \SI{15.41}{\giga\hertz} and \SI{15.92}{\giga\hertz}, yielding a frequency difference of $\Delta f \approx \SI{510}{\mega\hertz}$.
(b) Line scans of the dynamic magnetization projected along the central path of the Möbius strip, shown for several modes. The numbers above each scan indicate the number of wavelengths (in units of $2\pi$ phase) along the circular path.
The scans exhibit odd half-integer mode numbers, consistent with the antisymmetric boundary conditions imposed by the Möbius topology. 
 Clockwise and counterclockwise modes display visibly different wavelengths.
}
\end{figure}

Unlike the nearly degenerate mode doublets observed in the bent ring, the spin-wave modes in the Möbius strip appear in closely spaced pairs but with clearly distinct frequencies, cf.~Fig.~\ref{fig:doublets_mob}a). To understand this frequency splitting, it is helpful to consider the standing-wave modes in the bent ring as superpositions of two counterpropagating spin waves with identical frequency and wavelength. The resulting mode profiles differ only in their phase, i.e., in the positioning of nodes and antinodes around the ring—leading to mode doublets with nominally identical eigenfrequencies.

In the Möbius strip, this degeneracy is lifted, resulting in pairs of eigenmodes with clearly separated frequencies. The formation of frequency doublets has been predicted in the literature in the case of planar magnetic nanorings with vortex configurations under perpendicular magnetic fields, where the splitting arises from field-induced symmetry breaking \cite{ivanov_high_2005}. In contrast, the origin of the splitting in our case lies in the geometry: the continuous twist of the strip introduces a geometric phase~\cite{frank_wilczek_geometric_1989} (a Berry phase~\cite{berry_classical_1985} or Hannay angle~\cite{hannay_angle_1985}) that accumulates as the spin wave propagates along the structure. This geometric phase depends on the helicity of the strip and changes sign with the propagation direction. As a consequence, clockwise and counterclockwise spin waves experience different phase shifts over one complete circuit, which in turn modifies their effective wavelengths. Since the frequency of a spin-wave mode depends on its wavelength, this asymmetry lifts the degeneracy between the two propagation directions, resulting in the observed frequency splitting within each mode pair. As a result of this frequency splitting, a standing-wave mode, formed by the superposition of two counterpropagating waves of identical wavelength and frequency, is no longer possible in the Möbius strip.

In addition to the twist-induced Berry phase, spin-wave modes in the Möbius strip are constrained by quantization due to its non-orientable surface, which imposes antisymmetric boundary conditions. These conditions require the dynamic magnetization to change sign after one full circuit, leading to quantized modes with half-integer numbers of wavelengths along the strip's circumference. Figure~\ref{fig:doublets_mob}b) shows line scans of the dynamic magnetization component projected onto the local surface normal, taken along the strip’s central path. The scans reveal clear differences in wavelength between clockwise and counterclockwise modes within each doublet and confirm the appearance of odd-numbered half-integer wavelength modes, consistent with the Möbius geometry.

To disentangle the effects of topology-induced boundary conditions from those of the geometric phase introduced by the strip’s helical twist, we extended the analysis to double- and triple-twisted Möbius geometries. Circulating spin-wave modes also emerge in these structures, with only minor qualitative differences compared to the single-twist Möbius strip. As before, clockwise and counterclockwise propagating modes appear as doublets; however, the frequency splitting between them increases slightly with the number of twists. This trend suggests that the emergence of circulating modes is not dictated by the boundary conditions associated with non-orientability—which are restored to symmetry in the double-twist case—but rather by the presence and degree of helical twist in the strip geometry.

\subsection{Dispersion relations}
To deepen our understanding of spin-wave propagation in Möbius geometries, we analyze the dispersion relations of circulating spin-wave modes in single-, double-, and triple-twisted Möbius strips. Numerically extracting the wave vector for each eigenmode allows us to relate spin wave length and frequency and thereby to quantify how the geometry and twist affect spin-wave propagation.
\begin{figure}[ht]
    \centering
    \includegraphics[width=0.9\linewidth]{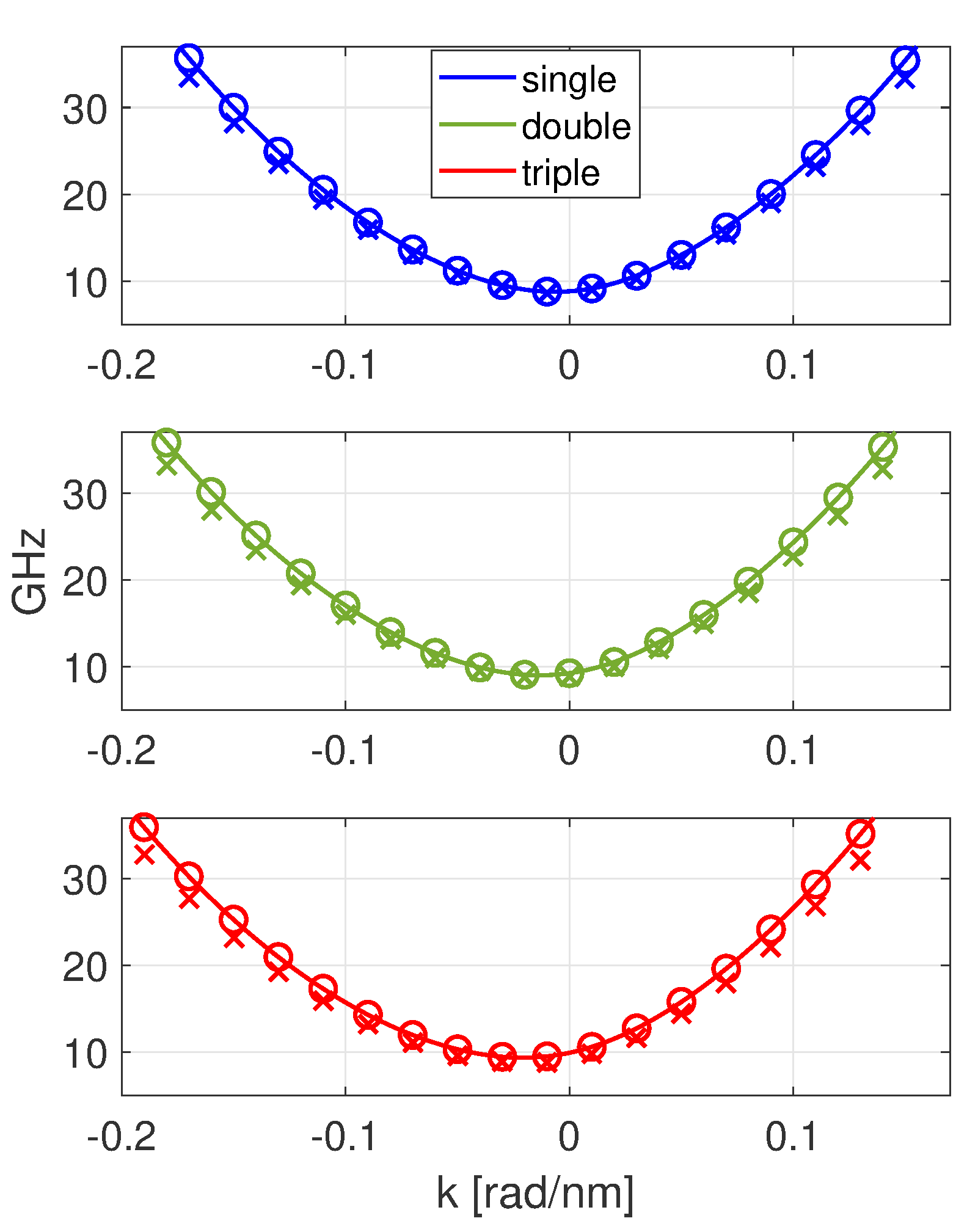}
    \caption{Comparison between theory and 3D FEM numerical results for
single, double, and triple left-handed Möbius strips. Solid lines refer to analytical dispersion relation \eqref{eq:dispersion curve}, open symbols correspond to quantization rule \eqref{eq:mobius quantization}, and ``cross'' symbols refer to FEM calculations. The values of parameters are:
$n=-1,-2,-3$, $R=50$ nm, $w=20$ nm, $d=2$ nm
 $N_1=0.1211$, $N_2=1-N_1$. The other material parameters are the same as in the previous figures.}
    \label{fig:cmp FEM analytical}
\end{figure}

The resulting dispersion relations, shown as cross symbols in Fig.~\ref{fig:cmp FEM analytical}, exhibit clear non-reciprocal behavior, with $\omega(k) \neq \omega(-k)$. This observation is consistent with previous analytical work on curved magnetic nanostrips~\cite{sheka_torsion-induced_2015, gaididei2017magnetization}, which predicted
non-reciprocal spin wave propagation due to geometric effects.
The dispersion curves display a distinct asymmetry, including off-centered minima that do not occur at $k = 0$. Here, positive and negative wavevectors correspond to clockwise and counterclockwise propagation, respectively. As the number of twists increases—from the single to the double- and triple-twisted Möbius strips—the asymmetry in the dispersion becomes more pronounced. This trend is consistent with the twist-induced geometric phase discussed earlier, which breaks the degeneracy between propagation directions and modifies the effective wavelength.

In order to provide further interpretation of the finite element calculations, we compare them with analytical formulas for spin wave dispersion relations in twisted nanostrips. The nanostrips have rectangular cross sections with thickness $d$ comparable to the exchange length $\ell_\mathrm{ex}$ of the material, their length $L$ is much larger than the width $w$, and the axis curve (middle line) has constant curvature $\kappa$ and torsion $\tau$.

The latter assumptions imply that the magnetization profile exhibits no change along the thickness and weak variation along the width, which means that the spatial dependence occurs mainly along the axis having a curvilinear abscissa $u\in [0, L]$. When the equilibrium magnetization is quasi-tangential to the strip axis, the spin wave oscillations at each abscissa $u$ will occur in the plane $(\bm e_1(u),\bm e_2(u))$ defined by the axis (in-plane) normal and (out-of-plane) binormal, respectively. Moreover, the small thickness $d\sim\ell_\mathrm{ex}$ allows for treating magnetostatics in local form~\cite{gioia1997micromagnetics}, which allows for introducing demagnetizing factors $N_1, N_2$ for the rectangular cross section~\cite{aharoni_demagnetizing_1998}.

A comprehensive analysis of spin wave propagation in twisted nanostrips (also including nutation due to magnetic inertia) has been performed in ref.~\cite{daquino_nonreciprocal_2025}. The analytical formulas for spin wave dispersion relations relevant to the present work are readily obtained by neglecting inertial effects:
\begin{align}  \omega_\pm(k)&=-2k\tau\ell_\mathrm{ex}^2\pm \sqrt{\omega_{1}(k)\omega_{2}(k)} \, \label{eq:dispersion curve}\\
\omega_{1}(k)&=h_0+N_1+\ell_\mathrm{ex}^2(k^2+\tau^2+\kappa^2) \,,\\\omega_{2}(k)&=h_0+N_2+\ell_\mathrm{ex}^2(k^2+\tau^2) \,,
\end{align}
where $k$ is the spin wave wavenumber and $h_0$ is the internal effective field at equilibrium along the strip axis.

It is apparent that the presence of a nonzero torsion $\tau$ produces a symmetry breaking that reflects into the non-reciprocal spin wave propagation and the asymmetric dispersion curve~\cite{sheka_torsion-induced_2015}, while the curvature $\kappa$ only produces a positive spectral shift in the frequency response. 

Moreover, it is possible to show~\cite{daquino_nonreciprocal_2025} that the spin wave propagation along the nanostrip produces at the abscissa $u$ a geometric (Berry) phase accumulation $\theta(u)$ that arises from a continuous rotation of the spin-wave polarization plane $(\bm e_1(u),\bm e_2(u))$ directly connected with the torsion:
\begin{equation}
    \theta(u)=\int_{0}^{u} \tau(u') du' + C \quad, \label{eq:twist angle}
\end{equation}
with $C$ being an arbitrary constant phase-shift. The chirality of the strip expressed by the sign of the torsion $\tau$ is responsible for the aforementioned lifted degeneracy of the spin wave doublets propagating in opposite directions.

The nontrivial topology of the M\"obius strip with curvature $\kappa=1/R$ ($R$ is the radius of the strip axis circle) and torsion $\tau=n/(2 R)$  ($n\in\mathbb{Z}$, the sign positive/negative implies right/left-handed chirality) induces a special quantization of the wavenumber due to anti-periodic boundary conditions~\cite{daquino_nonreciprocal_2025} that results in the following rule:
\begin{equation}
    k_h=\frac{n+2h}{2R} \quad,\quad h\in\mathbb{Z} \,. \label{eq:mobius quantization}
\end{equation}

The dispersion curves for single, double, and triple (i.e. $n=-1,-2,-3$ left-handed M\"obius strips) are reported in Fig.\ref{fig:cmp FEM analytical}. The excellent agreement with the theory is apparent both concerning the non-reciprocal character of the spin wave propagation and as well as the quantization rule \eqref{eq:mobius quantization} for wavenumbers induced by the nontrivial topology of the Möbius strips (see the open and 'cross' symbols in Fig.\ref{fig:cmp FEM analytical}).

\subsection{Forced oscillations\label{forced_osc}}
While the eigenmode analysis reveals the system's intrinsic high-frequency characteristics, a more experimentally relevant scenario involves the response to an external oscillating field. In such cases, a high-frequency (RF) field induces forced, steady-state oscillations, leading to characteristic absorption peaks at resonant frequencies.

\begin{figure}[ht]
\includegraphics[width=\linewidth]{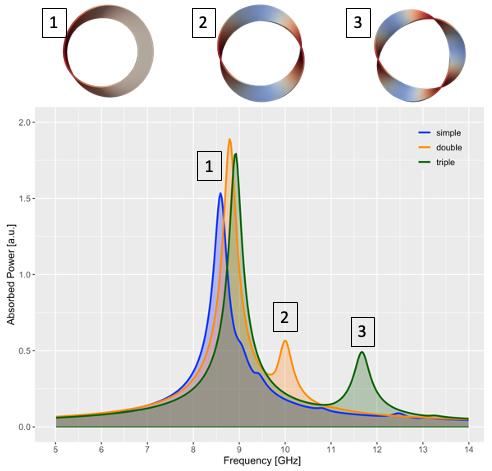}
\caption{Absorption spectra of the single-, double-, and triple-twisted Möbius strips under an externally applied oscillating magnetic field. The frequency range is chosen to highlight the dominant resonance peaks. The RF field is applied perpendicular to the ring plane (along the $z$ axis). The top panels show the dynamic magnetization profiles corresponding to the labeled peaks \fbox{1}, \fbox{2}, and \fbox{3} in the spectra below. Color indicates the $z$ component of the dynamic magnetization $\delta\bm{m}$, and the snapshots are taken at the phase where the average $z$-component $\langle\delta m_z\rangle$ reaches its maximum.
\label{fig:rf_spectra}}
\end{figure}

Fig.~\ref{fig:rf_spectra} shows the RF absorption spectra for the single-, double-, and triple-twisted Möbius strips. All three geometries exhibit a prominent resonance near \SI{9}{\giga\hertz}, with only minor shifts in frequency. This main absorption peak corresponds to the lowest-frequency eigenmode, located at \SI{8.60}{\giga\hertz}, \SI{8.79}{\giga\hertz}, and \SI{8.82}{\giga\hertz} for the single-, double-, and triple-twisted cases, respectively. This mode behaves as a ferromagnetic resonance (FMR), with the magnetization oscillating in phase across the entire structure, though with strongly inhomogeneous amplitude. The largest oscillation amplitudes occur in regions where the strip width is aligned along the $z$ axis. Because this mode lacks propagating character and involves uniform phase oscillation, the geometric and topological differences between the structures have little effect on its dynamics, resulting in nearly identical FMR responses across all three Möbius variants.

In addition to the principal FMR peak, the double- and triple-twisted Möbius strips exhibit secondary resonance peaks at \SI{9.9}{\giga\hertz} and \SI{11.67}{\giga\hertz}, respectively. In the double-twisted case, this secondary peak corresponds to a propagating spin-wave mode (eigenmode \#3, labeled as \fbox{2} in Fig.~\ref{fig:rf_spectra}) whose profile shows in-phase oscillations in the two high-susceptibility regions—where the strip width is aligned along the $z$ axis. A similar situation arises in the triple-twisted Möbius strip, which features three such areas. There, mode \#5 (labeled as \fbox{3} in Fig.~\ref{fig:rf_spectra}) represents a propagating wave whose wavelength allows constructive, in-phase oscillations across all three high-susceptibility zones. These resonances illustrate how the geometry selectively enhances modes that match the spatial distribution of magnetic susceptibility.

\begin{figure}[ht]
\includegraphics[width=\linewidth]{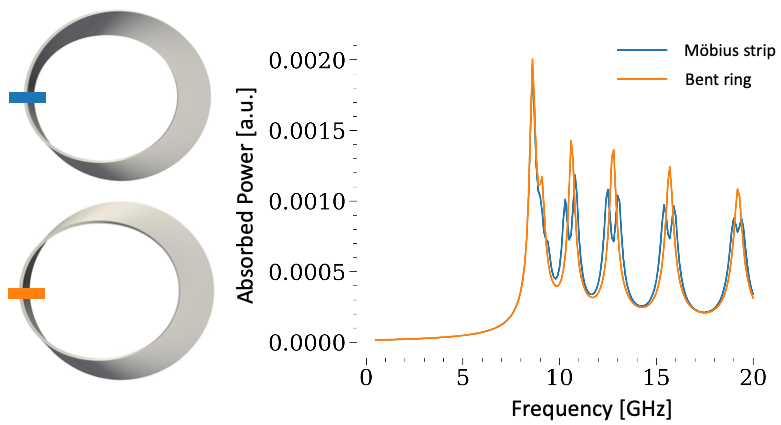}
\caption{Absorption spectra of the Möbius strip and the bent ring under localized RF excitation. The excitation field is applied to a small, high-susceptibility region in each structure, as indicated by the colored rectangles in the schematics on the left (blue for the Möbius strip, orange for the bent ring). In both cases, the RF field is oriented along the $z$ axis, perpendicular to the ring plane. The bent ring exhibits single-peak resonances at well-defined frequencies. In contrast, the Möbius strip displays double-peak structures with closely spaced frequencies, reflecting the non-degenerate nature of the counterpropagating spin-wave modes.
\label{fig:local_forced_mob_triv}}
\end{figure}

However, aside from these geometry-matched cases, a globally applied RF field with uniform phase is generally ineffective at exciting the well-defined propagating spin-wave modes discussed in Section~\ref{eigenmodes}. To selectively excite such modes, we apply a localized RF field restricted to a specific region of the strip. This approach allows individual eigenmodes to be excited by tuning the driving frequency to match their resonance. An example of this mode-selective excitation is shown in Fig.~\ref{fig:local_forced_mob_triv}, which presents the absorption spectra of the Möbius strip and the bent ring under time-harmonic excitation localized in the high-susceptibility region—where the strip is oriented such that its width lies along the $z$ axis.

In the topologically trivial bent ring, we observe sharp resonance peaks at well-defined frequencies, each corresponding to one of the nearly degenerate standing-wave mode doublets. Because the excitation is spatially localized, only a subset of modes is activated, specifically those for which the field is applied at or near an oscillation antinode. In our example, these correspond to the bottom row of modes shown in Fig.~\ref{fig:modes_bent}. In contrast, the Möbius strips exhibit resonance doublets with closely spaced frequencies, consistent with the eigenmode structure discussed in Section~\ref{eigenmodes}, where each doublet corresponds to a pair of counterpropagating spin-wave modes. The use of a localized, linearly polarized RF field thus enables the excitation of spin waves with well-defined frequency and propagation direction, effectively selecting between clockwise and counterclockwise modes.
\section{Discussion}
Our results demonstrate how geometric twist and nontrivial topology fundamentally alter spin-wave dynamics in three-dimensional magnetic nanostructures. Using Möbius-shaped soft-magnetic strips, we show that the interplay between helical curvature and non-orientable topology leads to symmetry breaking in spin-wave propagation. In contrast to the topologically trivial bent ring, which supports degenerate standing-wave modes, Möbius strips exhibit frequency-split doublets corresponding to counterpropagating spin-wave modes. While the Möbius strip is non-orientable, we find that it is not the topology that drives this effect, but rather the presence of a net geometric twist along the propagation path. The bent ring, though curved, features twist-compensation between its two halves, resulting in an effective zero net twist and, correspondingly, no Berry-phase-induced splitting. In the Möbius geometries, the continuous helical twist introduces a geometric (Berry) phase that modifies the effective wavelength depending on direction, breaking propagation symmetry and lifting the degeneracy between clockwise and counterclockwise modes.
 As a result, the Möbius strips support directionally selective, non-reciprocal spin-wave modes that cannot form standing waves. These effects grow more pronounced with increasing twist, as confirmed by our analysis of double- and triple-twisted Möbius geometries.

 The non-orientable topology of the Möbius strip imposes antisymmetric boundary conditions, which result in half-integer wavelength quantization. Spatially localized RF excitation enables controlled excitation of specific eigenmodes and selection of propagation direction—a mechanism that could be extended to programmable magnonic routing or phase-encoded information transfer.

Our analytical model, which closely matches the numerical dispersion data, provides a predictive framework for designing twisted magnonic systems and interpreting their spectral features.

These findings highlight the potential of using geometric and topological design principles to engineer magnonic responses in 3D systems. Future work may explore how these effects evolve in interacting Möbius arrays, or systems with Dzyaloshinskii–Moriya interaction~\cite{volkov_mesoscale_2018}. Our approach also opens up opportunities for integrating non-orientable magnetic structures into on-chip spintronic architectures where control over spin-wave phase, directionality, or non-reciprocity is desirable.
\section{Methods}

\subsection{Mesh generation and static magnetization structure}
Three-dimensional models of the Möbius-type geometries and the bent ring were generated using the open-source software FreeCAD~\cite{FreeCAD}. Tetrahedral finite-element meshes were produced with NETGEN~\cite{NETGEN}, using a target cell size of \SI{1}{\nano\meter}. This discretization resulted in problem sizes of slightly more than \num{50000} elements for each geometry.
The zero-field static magnetization configurations were computed with our GPU-accelerated open-source code \texttt{tetmag}~\cite{hertel_tetmag_2023}.

\subsection{Eigenmodes and forced oscillations}
Eigenmodes were computed using a dedicated algorithm described in Ref.~\cite{daquino_micromagnetic_2023}, which implements the formalism originally developed by d'Aquino et al.~\cite{daquino_novel_2009}. The method is based on a linearized form of the Landau–Lifshitz–Gilbert equation.
Initially designed for the calculation of eigenmodes, the approach was later generalized~\cite{daquino_micromagnetic_2023} to solve micromagnetic problems involving stationary magnetic oscillations in an applied RF field.
For the forced-oscillation calculations discussed in Section~\ref{forced_osc}, we assumed a Gilbert damping constant of $\alpha = 0.01$ and an RF field amplitude of \SI{0.5}{\milli\tesla}.

\subsection{Numerical extraction of dispersion relations}
Unlike extended magnetic waveguides, Möbius-type geometries form closed loops, which prevents the direct application of standard techniques for extracting dispersion relations. To address this, we map the three-dimensional eigenmode profiles onto an effective one-dimensional coordinate that traces the central path of the strip. This procedure conceptually ``unrolls'' the Möbius geometry into a helical waveguide, rendering it amenable to spatial Fourier analysis.
By performing a Fourier transform of the dynamic magnetization along this path, we identify the dominant wavevectors associated with each eigenmode and thereby construct the spin-wave dispersion relation. The $k$-values obtained from this analysis correspond to the spin-wave wavelengths observed in the line scans shown in Fig.~\ref{fig:doublets_mob}b.

\section*{Acknowledgments}
This work was funded by the France 2030 government investment plan managed by the French National Research Agency ANR under grant reference PEPR SPIN – [SPINTHEORY] ANR-22-EXSP-0009.
The authors acknowledge the High Performance Computing Center of the University of Strasbourg for supporting this work by providing access to computing resources. M.d'A. acknowledges support from the Italian Ministry of University and Research, PRIN2020 funding program,
grant number 2020PY8KTC. 


\end{document}